\documentclass[a4paper,twoside]{article}

\usepackage{epsfig}
\usepackage{subcaption}
\usepackage{calc}
\usepackage{amssymb}
\usepackage{amstext}
\usepackage{amsmath}
\usepackage{amsthm}
\usepackage{multicol}
\usepackage{pslatex}
\usepackage{apalike}
\usepackage[bottom]{footmisc}
\usepackage[hidelinks]{hyperref}
\usepackage[nolist]{acronym}
\usepackage{xcolor}
\usepackage{SCITEPRESS} % Please add other packages that you may need BEFORE the SCITEPRESS.sty package.

\begin{acronym}
\acro{nlp}[NLP]{natural language processing}
\acro{lm}[LM]{Language Model}
\acro{plm}[PLM]{Pretrained Language Model}
\acroplural{plm}[PLMs]{Pretrained Language Models}
\acro{kg}[KG]{knowledge graph}
\acroplural{kg}[KGs]{knowledge graphs}
\acro{p}[P]{Precision}
\acro{r}[R]{Recall}
\acro{mrr}[MRR]{Mean Reciprocal Rank}
\acro{mag}[MAG]{Microsoft Academic Graph}
\acro{fos}[FoS]{field-of-study}
\acroplural{fos}[FoS]{fields-of-study}
\acro{rims}[RIMS]{Research Information Management System}
\acroplural{rims}[RIMS]{Research Information Management Systems}
\end{acronym}

\begin{document}

\title{A Knowledge Graph Approach for Exploratory Search \\ in Research Institutions}

% authors for INSTICC version
%\author{\authorname{Tim Schopf\orcidAuthor{0000-0003-3849-0394}, Nektarios Machner\orcidAuthor{0009-0001-8359-6668} and Florian Matthes\orcidAuthor{0000-0002-6667-5452}} \affiliation{Department of Computer Science, Technical University of Munich, Germany}\email{\{tim.schopf, nektarios.machner, matthes\}@tum.de}}  

% authors for arXiv submission 
\author{\authorname{Tim Schopf, Nektarios Machner and Florian Matthes} \affiliation{Department of Computer Science, Technical University of Munich, Germany}\email{\{tim.schopf, nektarios.machner, matthes\}@tum.de}}

%\author{\textbf{Anonymous submission}}

\keywords{Knowledge Graphs, Exploratory Search}

\abstract{Over the past decades, research institutions have grown increasingly and consequently also their research output. This poses a significant challenge for researchers seeking to understand the research landscape of an institution. The process of exploring the research landscape of institutions has a vague information need, no precise goal, and is open-ended. Current applications are not designed to fulfill the requirements for exploratory search in research institutions. In this paper, we analyze exploratory search in research institutions and propose a knowledge graph-based approach to enhance this process.}

\onecolumn \maketitle \normalsize \setcounter{footnote}{0} \vfill

\section{\uppercase{Introduction}}
\label{sec:introduction}

Scientific literature has grown exponentially over the past centuries, with a two-fold increase every 12 years \cite{10.1145/3097983.3098016}. Concurrently, the number of research institutions as well as the number of researchers and research areas within these institutions has also been growing. Once research institutions reach a certain size, it becomes challenging to determine which topics are being researched at an institution and who is researching which topics. The most basic approach to disclosing ongoing research at research institutions is to post this information as unstructured text on the institution's or its sub-units' websites. Usually, these websites are designed according to the organizational structures of research institutions rather than research areas, further complicating the process of understanding what research is being conducted. More advanced solutions attempt to consolidate the entire research output of researchers and research units in an institution using specific \acp{rims}. These \acp{rims} can identify and visualize the domain of expertise of researchers and research units based on research topic tags automatically extracted from publications. This may accelerate the process of a targeted search, such as finding an expert in a specific domain. However, these systems are not capable of representing the relationships between individual research areas and assess the similarity of researchers based on organizational affiliations. As a result, the search for related research areas and further potentially relevant experts remains challenging. In addition, \acp{rims} often lack comprehensible statistics and analyses about the specific research fields of researchers and research units. Therefore, despite RIMS, it is still very time-consuming for researchers to obtain an overview of the research landscape of a research institution. 

To understand the current process of researchers seeking insights into the research landscape of institutions, we conducted several interviews. Our analysis is based primarily on one-on-one interviews with researchers, ranging from early to late career stages. From this, we conclude that the use of \acp{rims} is still not widely adopted. Consequently, researchers currently heavily rely on search engines to find relevant institution websites and then use these as a starting point for both browsing and formulating new search queries. The process of searching and browsing continues iteratively until researchers have a satisfactory overview of an institution's research landscape. However, simply using returned lists of relevant elements that need to be analyzed manually limits the knowledge search of researchers \cite{10.1145/3360901.3364435,brainard2020scientists}. Two main reasons can be identified for this. First, it is very tedious and time-consuming to search and browse all relevant websites. Second, important information that is potentially relevant to researchers but was not searched for because the appropriate search queries are unknown can be missed. 

We argue that \textit{the exploratory knowledge search process for research activities in research institutions can be significantly enhanced by the semantic linking of research topics and their subsequent association with further relevant entities.} %using a research topic-based user navigation and thereby enable the semantic linking of relevant entities.

Our aim is to advocate for more clarity and conciseness in presenting the areas of competence of research institutions. In addition, we propose an approach to enhance the exploratory knowledge search process for research activities in research institutions. %\textcolor{red}{It is based on a \ac{kg} powered application that can adequately represent the semantic associations between different relevant entities.} 
Researchers seeking an overview of current research at an institution should be able to identify potential collaborators more easily and be less likely to miss important information. Enhancing this knowledge search process has the potential to encourage and facilitate more research collaborations within research institutions. Eventually, enhanced knowledge search and the resulting potential for more research collaborations may also have a positive impact on an institution's overall research output.

\section{\uppercase{Requirements Analysis}}
\label{sec:requirements-analysis}

\subsection{Exploratory Search in Research Institutions}

The information need for obtaining an overview of an institution's research landscape is complex, evolves during the search, and is open-ended. In this case, the user objective is rather vague and can be divided into several smaller sub-goals during the search (e.g., which research areas exist at the institution, who works in certain research areas, or what exactly is being studied in the research areas). Furthermore, users can not rely on a single search result to satisfy their information need. They need to assemble the relevant information from multiple search results and institution websites to get an overview of the research landscape. Users perform several reformulations and refinements of their search queries based on the new information obtained from institution websites. The search process continues until users consider they have obtained enough information to reach their vague goal. However, users may never know if their acquired knowledge is complete or if they have missed important aspects of the research landscape at the institution. Therefore, the search does not have a defined end, but can be continued after receiving additional information, if users feel that there are still unanswered questions to be investigated. 

Existing search engines are optimized for simple lookup searches, which are characterized by low complexity and a precise objective \cite{https://doi.org/10.1002/asi.23617}. However, the described knowledge search for research activities in research institutions has a rather vague, complex, and open-ended goal, showing many characteristics typical of an exploratory search \cite{10.1145/3038462.3038465,grether_and_witschel_kdir22}. In our interviews, users were able to use search engines to quickly find answers to specific questions, such as whether there exists a particular research area at a research institution. However, in the case of rather vague objectives, e.g., searching for key research areas of a particular department in an institution, the search process turned out to be quite inefficient and took a considerable amount of time. Moreover, users were not able to properly assess whether their results were exhaustive or incomplete at the end of the search process. We conclude that it is not sufficient to rely only on existing search engines to gain insights into the research landscape of institutions. Moreover, we argue that a new approach is required to sufficiently address the need for exploratory search in research institutions.

\subsection{Requirements}

To investigate the exploratory search process of researchers seeking insights into the research landscape of institutions, we conducted semi-structured interviews. Our participants included graduate students, PhD students, and professors. We gave each participant an exploratory task with a rather vague objective, such as e.g., identifying the most relevant research areas at an unknown research institution. They were allowed to complete this task using any resources at their disposal. In many cases, search engines were the tool of choice. We observed the search and asked participants to express their thought processes aloud. Throughout the search, we kept asking questions about the challenges participants encountered and what might help them to better deal with them. 

From our interviews, we derive the following key requirements for an application designed to enhance the exploratory knowledge search process for research activities in research institutions: \newline

\begin{itemize}

    \item \textbf{Requirement 1: Hierarchical navigation structure based on research topics}

    Users are used to navigating through applications and websites in a hierarchical manner. Thereby, more general concepts are presented in the upper levels and more specific concepts are presented in the lower levels of the navigation hierarchy. An interface using hierarchical navigation simultaneously shows previews of where to go next and how to return to previous states in the exploration \cite{hearst-stoica-2009-nlp}. In the scientific domain, the most important concepts are the specific research topics and areas. Therefore, to enable a streamlined and satisfactory semantic exploration of scientific knowledge, we need to use a well-organized hierarchical structure of research topics and areas that comprehensively covers the broad spectrum of academic disciplines \cite{shen-etal-2018-web}. \newline
    
    \item \textbf{Requirement 2: Semantic linking of scientific entities}

    In addition to the hierarchically structured research topics, the application has to include accurate representations of the research institution and its research units as well as the associated researchers and their publications. The respective entities have to be semantically linked to ensure an extensive exploratory search that offers many opportunities for discovering new information. Most importantly, the relationships between the hierarchically structured research topics and the other entities must be modeled precisely. This allows for a comprehensive exploratory search within the research institution based on research topics. In addition, other relationships such as the affiliation of researchers with research units or the similarity of researchers need to be modeled in order to recommend other relevant entities during the search process. \newline

    \item \textbf{Requirement 3: Single Source of Truth}

    As a countermeasure to the complex and open-ended nature of exploratory search, the application has to act as a single source of truth containing all relevant data of one research institution. With all relevant data being available from one source, the user should feel confident in receiving a comprehensive overview and it should not be necessary to involve further external tools and search engines to find the desired information. \newline
    
    \item \textbf{Requirement 4: Support for lookup search as well as recommendations}

    Exploratory search involves a significant amount of lookup activities in addition to a wide range of other goals and tasks \cite{White2009}. Therefore, the application has to implement a semantic search that supports basic lookup tasks. Furthermore, based on the semantic relationships, the application must be able to recommend further relevant entities for the user to explore. This facilitates the discovery of new and relevant knowledge that was potentially unknown to users prior to conducting the exploratory search. \newline

    \item \textbf{Requirement 5: Aggregation of data points into insights}
    
    The application should provide both a static overview of all relevant data, as well as dynamic aggregations of important data points displayed in meaningful ways, in order to facilitate drawing impactful conclusions for the respective research fields.

\end{itemize}

\begin{figure*}[ht!]
    \centering
    \includegraphics[width=\textwidth]{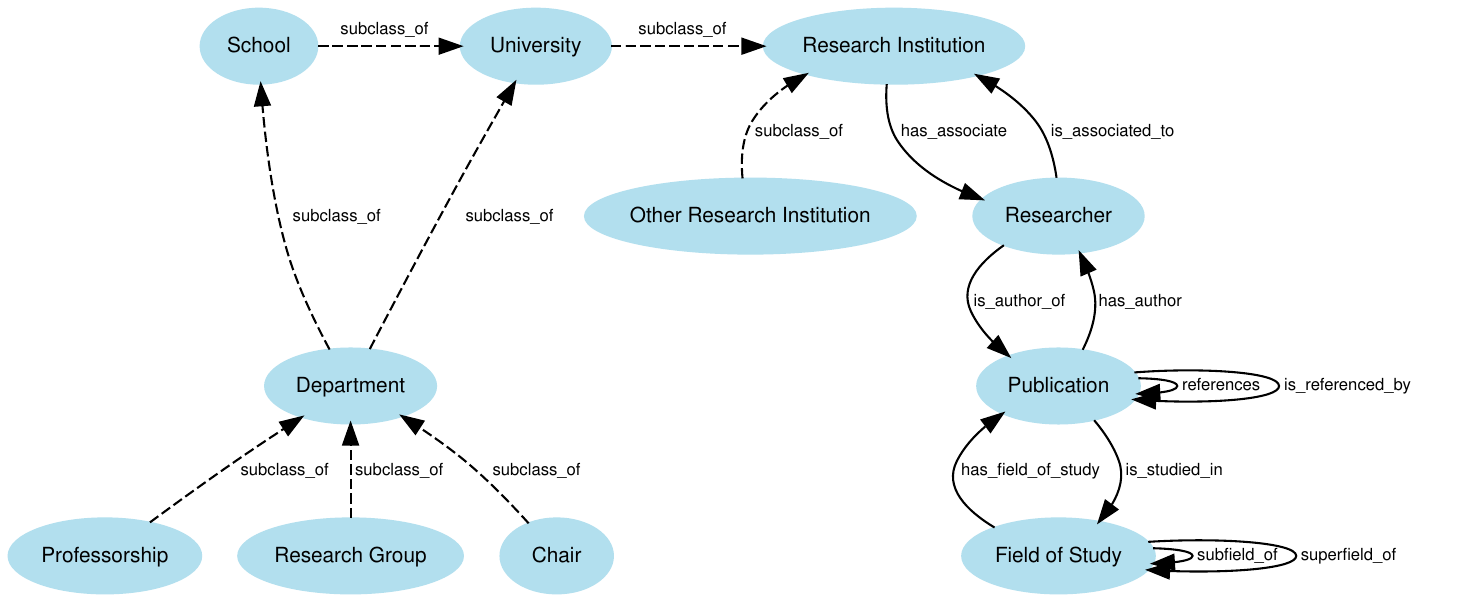}
    \caption{Simplified ontology of the research institution knowledge graph. Here, only the \textit{University} entity, representing German universities, with its respective sub-units is expanded. Other types of research entities are grouped in the node \textit{Other Research Institution}. All entities are transitively connected to the \textit{Field of Study} entity. This means that each researcher and each sub-unit of the research institute can be linked to their respective \acl{fos} by traversing the \acl{kg}.}
    \label{fig:ontology}
\end{figure*}

\section{\uppercase{Constructing the Research Institution Knowledge Graph}}
\label{sec:constructing}

In recent years, \acp{kg} have established as an approach for semantically representing knowledge about real-world entities in a machine-readable format \cite{schneider-etal-2022-decade}. In contrast to relational databases or document databases, \acp{kg} can explicitly capture all kinds of semantic relationships between different entities. Since our data is highly interconnected and the primary focus of this approach is on data retrieval and analysis, we propose the usage of a \ac{kg} as the database for exploratory search in research institutions. 

To construct the \ac{kg}, first a hierarchy of research topics is needed that can be used to semantically link other relevant entities. Manually defined hierarchies although very precise, are usually very domain-specific or generic and cannot fully capture the whole range of existing research topics in a research institution. Therefore, we propose to use the automatically constructed \ac{fos} hierarchy of the \ac{mag} \cite{10.1145/2740908.2742839,shen-etal-2018-web,10.1162/qss_a_00021}. The hierarchy consists of over 200K hierarchically structured \ac{fos} concepts, covering a broad spectrum of academic disciplines. Thereby, each \ac{fos} concept represents one distinct academic discipline. After obtaining the \ac{fos} hierarchy, the other entities need to be semantically linked to specific research topics. To this end, the publications of researchers can be classified according to the existing \ac{fos} concepts. Classification of articles according to their related \ac{fos} concepts can be performed by using semantic similarity scores between the respective text representations of concepts and publications \cite{shen-etal-2018-web,toney-dunham-2022-multi}. Subsequently, the ontology shown in Figure \ref{fig:ontology} can be used as a data model to semantically link the remaining entities. Finally, the classification of publications together with transitive relations can be used to infer the \ac{fos} concepts of researchers and their associated sub-units.

\section{\uppercase{Proof of Concept}}

\begin{figure*}[ht!]
    \centering
    \includegraphics[width=\textwidth]{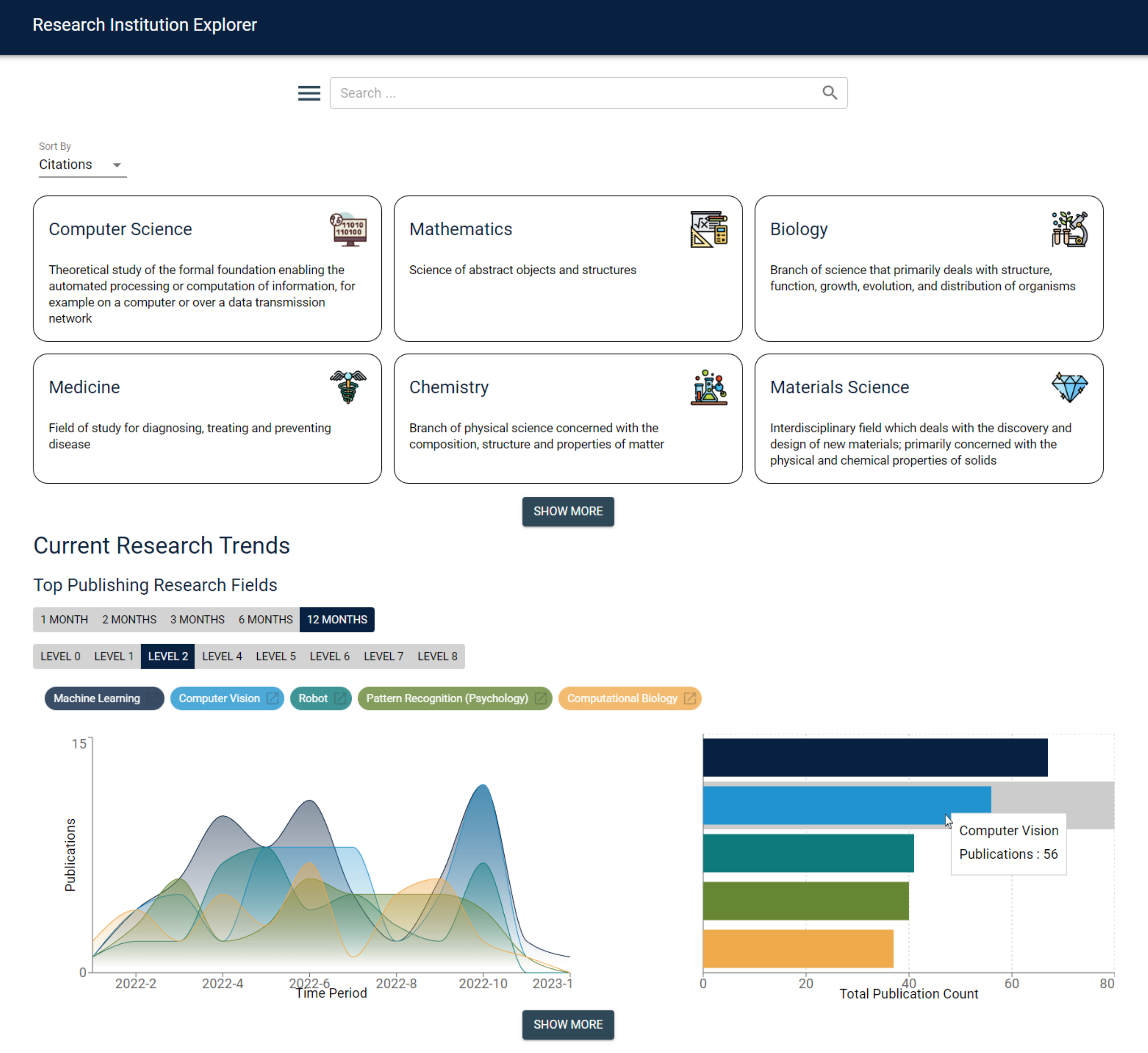}
    \caption{Mock-up of the proof of concept homepage. Users can either use the search to retrieve specific information or explore the existing research landscape at the institution by using the hierarchically structured \ac{fos} concepts. On the homepage, the \ac{fos} concept of each high-level research domain is shown in a convenient layout.  Additionally, users can immediately gain insights into current research trends within the institution. Here, users can examine research trends within different depth levels of the \ac{fos} hierarchy.}
    \label{fig:mockup}
\end{figure*}

\label{sec:features}
\subsection{Architecture}

We propose a proof of concept based on a classic client-server architecture. Thereby, the backend consists of a NodeJS app which operates as a server and is connected to a Neo4j database containing all semantically linked data. The frontend is a client-side single-page application implemented in React with a thin server architecture where most business logic is moved from the server to the client that requests data only as needed, thereby allowing for a seamless user experience.

\subsection{Data Model}

As data model of the application, we propose the ontology shown in Figure \ref{fig:ontology}. For a prototypical implementation \textit{Research Institution} should the central entity. This entity concept, which can be substituted by several sub-classes, is linked to important entities such as researchers and transitively also to publications, as well as to \aclp{fos}. The data model can be discretionarily extended to support further entities and their associated data.

\subsection{Features}

\begin{itemize}
    \item \textbf{Search \& Browse}
    
    Users can start looking for the desired information either by using the full-text search to search for specific data entities from the underlying model, or they can browse a sortable list of research fields that serves as a starting point to navigate deeper into the data hierarchy and explore the related data by following the embedded links. As shown in Figure \ref{fig:mockup}, users can use the input field to search in the application and browse through research fields using the \ac{fos} tiles.
    \newline
    
    \item \textbf{Statistics \& Analytics}
    
    All data entities are enriched with meaningful statistics and analytics supported by graph visualizations for a better understanding. Among other things, it is possible to compare research fields regarding amount of citations or check which research topics are currently trending. Figure \ref{fig:mockup} illustrates how the data can be used to display current research trends within the institution. \newline
    
    \item \textbf{Semantically linked data}
    
    Since all data is semantically linked, it is possible to navigate to related content from any point in the data hierarchy. Furthermore, the application is able to identify and recommend additional related content by means of a similarity search that is automatically conducted over the data set, thereby complementing the identification of related content through metadata.
\end{itemize}

\section{\uppercase{Limitations}}
\label{sec:limitations}
%State some limitations but conclude that this approach is still the best. -> Support and acknowledge the opposing points. Just be sure you aren't discrediting your own views. Explain that your position is still the best one, despite the strength of counter-arguments. This is where you can work to discredit some of the counter-arguments and support your own. 
Since our proposed solution requires a single source of truth, it is vital that relevant data is available and of high quality. Complex data structures are prone to becoming stale quickly, which reduces the overall expressiveness of the presented information. The necessity to acquire all relevant data as well as to keep it up-to-date and consistent requires the maintainers of the data source to continually check and if necessary update the data. 
    
By design, the solution is limited to all relevant data of one research institution. Further enriching the data hierarchy with semantic links to external sources is out of scope. Additionally, since all \ac{fos} concepts are inevitably tied to publications, only those research areas for which publications exist can be represented. 

Due to the inherent complexity of displaying large amounts of data in a clear and concise manner, our proposed approach still requires users to invest time before they can gain an extensive overview of more complex topics.

\section{\uppercase{Conclusion}}
\label{sec:conclusion}

Understanding the landscape of research institutions is a challenging task for researchers. Current solutions do not fulfill the requirements imposed by the resulting exploratory search process. Through several interviews we derived key requirements for a possible solution that we prototypically sketched out as a minimal web application serving as proof of concept. Our proposed application is able to provide extensive information for a rather vague exploratory objective. Further, it enables users to gain insights into the desired search space by semantically linking related entities and aggregating relevant data points into insightful analytical views. However, the approach is limited by the quality and availability of relevant data and is restricted to data of one research institution. Since accumulating large amounts of data and presenting them in an exhaustive yet concise manner is inherently complex, the degree to which such a task can be simplified is limited and still requires the user to invest a certain amount of time and effort in order to reach the desired goal. That being said, we do believe that having relevant data semantically linked and easily accessible within one application helps alleviate the necessary time and effort investment to achieve an exhaustive overview of a particular research landscape. In future work, we aim to implement the proposed application as a minimum viable product that we can then systematically evaluate. To further enhance the proposed approach, we subsequently plan to extend the initial implementation based on the user feedback.

\bibliographystyle{apalike}
{\small
\bibliography{custom,anthology}}

\end{document}